\documentclass[aps,pra,twocolumn,superscriptaddress]{revtex4-1}

\usepackage{soul}
\usepackage{color}
\usepackage{amsmath}
\usepackage{graphicx}
\usepackage{epstopdf}
\usepackage{mathtools}
\usepackage{natbib}
\begin{document}
	\title{Exceptional points-based optical amplifiers}
	\author{Q. Zhong}
		\affiliation{Department of Physics and Henes Center for Quantum Phenomena, Michigan Technological University, Houghton, Michigan, 49931, USA}
	
	\author{S.K. Ozdemir }
		\affiliation{Department of Engineering Science and Mechanics, The Pennsylvania State University, University Park, Pennsylvania 16802, USA}

	\author{A. Eisfeld}
	\affiliation{Max Planck Institute for the Physics of Complex Systems, N\"othnitzer Strasse 38, 01187 Dresden, Germany}
	
	\author{A. Metelmann}
	\email[Corresponding author: ]{a.metelmann@fu-berlin.de }
	\affiliation{Dahlem Center for Complex Quantum Systems and Fachbereich Physik, Freie Universit\"{a}t Berlin, 14195 Berlin, Germany}

	\author{R. El-Ganainy}
	\email[Corresponding author: ]{ganainy@mtu.edu}
	\affiliation{Department of Physics and Henes Center for Quantum Phenomena, Michigan Technological University, Houghton, Michigan, 49931, USA}
    \affiliation{Department of Electrical and Computer Engineering, Michigan Technological University, Houghton, Michigan, 49931, USA}

\begin{abstract}
The finite gain-bandwidth product is a fundamental figure of merit that restricts the operation of standard optical amplifiers. In microcavity setups, this becomes a serious problem due to the narrow bandwidth of the device. Here we introduce a new design paradigm based on exceptional points, that relaxes this limitation and allows for building a new generation of optical amplifiers that exhibits better gain-bandwidth scaling relations. Importantly, our results can be extended to other physical systems such as acoustics and microwaves.
	
\end{abstract}
\maketitle

\section{Introduction}
The quest for new photonic devices and functionalities is currently pushing the limit for novel design paradigms and material platforms. One of the most fundamental processes in optical science and engineering is signal amplification. Current amplification mechanisms include incoherent pumping (atomic or band inversion followed by stimulated emission) or coherent pumping (such as in nonlinear wave mixing processes). Based on their geometry, semiconductors optical amplifiers (OAs) \cite{Dutta-SOA,Connelly-SOA} can be classified into traveling \cite{Hattori1994EL} or standing \cite{Amarnath2005PTL} waves devices. The former offers a larger bandwidth of operation at the expense of the attainable gain values and footprint (few millimeters in length). On the other hand, the latter can have larger gain due to the power recycling in the resonator which allows for a much smaller device size, suitable for large scale integration. However, the same resonant condition leads to a very narrow bandwidth. This fundamental limitation pertinent to cavity-based optical amplifiers (and generally also electronic and microwave amplifiers) is known as the gain-bandwidth product and is often expressed as: $\chi=\sqrt{G} \cdot B$ = const. \cite{Clerk2010RMP}, where $G$ is the maximum gain and $B$ is the bandwidth (which is usually defined as full width at half maximum (FWHM) of the power gain curve --- here we adopt this definition). Relaxing this constraint beyond its standard scaling will enable a new level of integration for high-performance photonic circuits. Going beyond the standard gain-bandwidth limit has been studied in parametrically driven coupled-mode systems \cite{Metelmann2014PRL, Metelmann2015PRX, Metelmann2017PRA, Ockeloen2016PRX, Hatridge2019arXiv}. 
There one combines parametric amplification with frequency conversion processes, which effectively removes the instability introduced by the amplification process.
Such multi-tone setups require well controlled pump-amplitudes and demand strong external driving, which can be rather challenging for the operation of the amplifier in terms of its stability.
Thus it would be desirable to develop simpler designs which exhibit improved gain-bandwidth behavior.

Here we introduce a new OA scheme based on optical resonators operating at exceptional points (EPs) -- a special type of singularities that arise in non-Hermitian Hamiltonians when two or more eigenstates coalesce \cite{Heiss2004JPA,Muller2008JPA,El-Ganainy2018NP,Feng2017NP}. We show that the gain-bandwidth product of the proposed device scales differently from that of standard resonators, which leads to superior performance without requiring any additional control tones. 
These predictions are confirmed by performing full-wave finite-difference time-domain (FDTD) analysis using realistic microring resonator geometries and material parameters.  

\begin{figure}[htbp]
	\centering
	\includegraphics[width=2.6in]{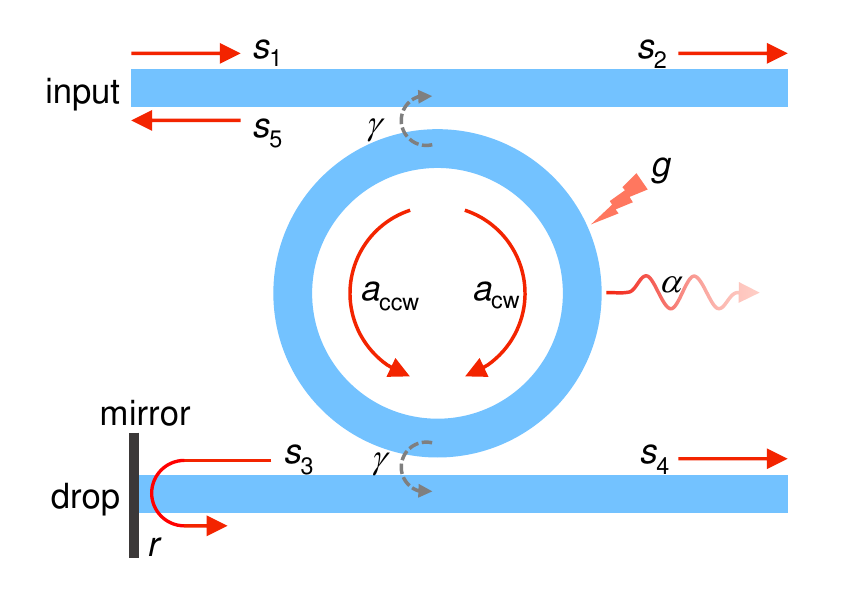}
	\caption{Schematic structure of an optical amplifier (OA) based on microring resonator working at an exceptional point (EP). The input $s_1$ will couple into the microring resonator (coupling rate $\gamma$) and be amplified by the pumping gain $g$. The clockwise mode $a_\text{cw}$ will couple into counterclockwise mode $a_\text{ccw}$  while the opposite is not true because of the mirror at the drop port. The output $s_5$ will be amplified in this process. Here $r$ is the magnitude of the field reflection coefficient of mirror and $\alpha$ is the decay rate due to radiation and material loss. }
	\label{Fig:Setup} 
\end{figure}

To this end, we consider the structure shown in Fig. \ref{Fig:Setup}. It consists of a microring resonator coupled to two identical waveguides, one of which is terminated by a mirror and the other is used as an input/output port. Optical gain is applied to the ring where the amplification process takes place. In the absence of the mirror, the system has two independent eigenmodes with identical resonant frequencies $\omega_0$: clockwise (CW) and counterclockwise (CCW); i.e. it operates at a diabolic point (DP). Under this condition and by using temporal coupled mode theory (TCMT) \cite{Vahala-OM,Fan2003JOSA}, we find the scattering coefficient between the input ($s_1$) and output ($s_3$) ports: 
\begin{equation}\label{Eq:s3}
	s_{31} \equiv \frac{s_3}{s_1}=\frac{2\gamma}{i(\omega-\omega_0)+2\gamma+\alpha-g},
\end{equation}
where $\alpha$ is the decay rate due to loss (radiation and material loss excluding those caused by coupling to the waveguides); $\gamma$ is the loss rate due to coupling to each of the two waveguides; $g$ is the applied gain rate and $\omega$ is the input signal angular frequency. From Eq. (\ref{Eq:s3}), we obtain the following expressions for maximum power amplification at resonance $G_\text{DP}\equiv|s_{31}(\omega_0)|^2=4\gamma^2/(2\gamma+\alpha-g)^2$,
and the bandwidth (in terms of angular frequency):   $B_\text{DP}=2(2\gamma+\alpha-g)$.
The gain-bandwidth product can then be expressed as $\chi_{\text{DP}}=\sqrt{G_\text{DP}}\cdot B_\text{DP}= 4\gamma$. The subscript $\text{DP}$ here emphasizes that these quantities are obtained for an OA operating at a DP.

\section{Amplification at exceptional points}
\begin{figure*}[htbp]
	\centering
	\includegraphics[width=5.35in]{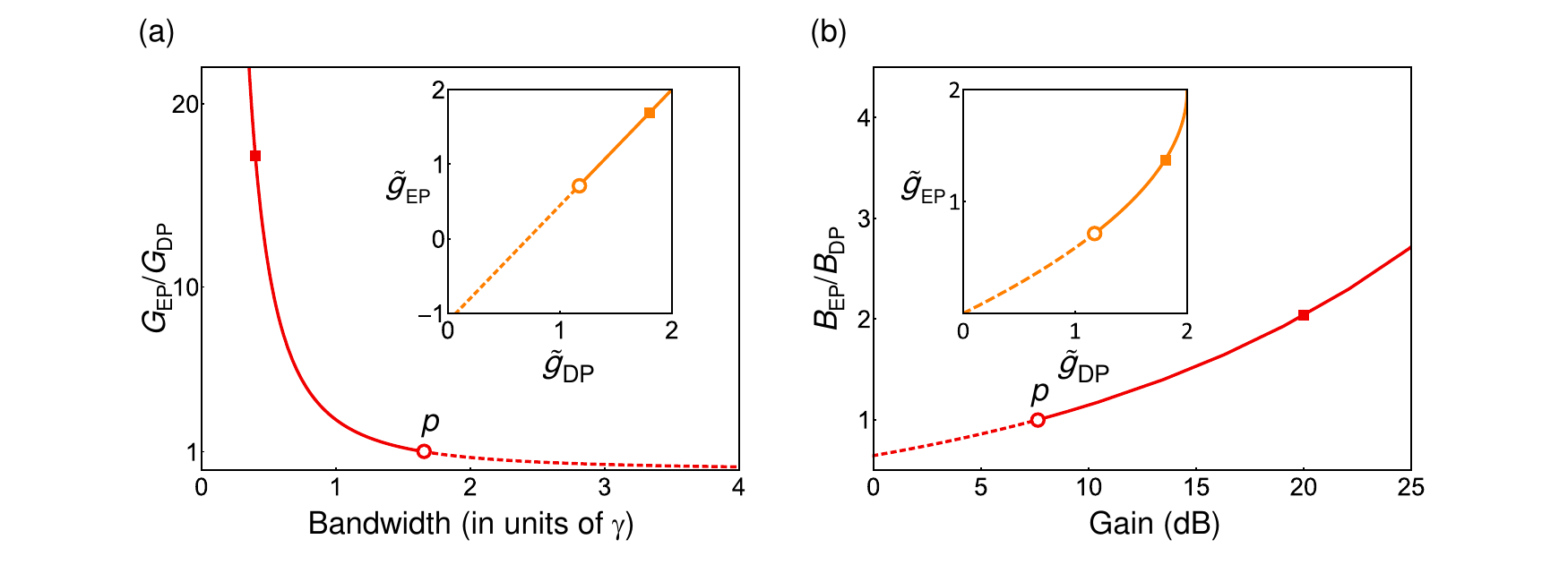}
	\caption{(a) Amplification enhancement for EP-based OA (as compared with standard DP-based resonators) as a function of their identical bandwidth as measured in units of $\gamma$. (b) Same as in (a) but for bandwidth enhancement as a function of the identical amplification. The insets in (a) and (b) plot the relations between the material gain values that are necessary to achieve identical bandwidth or amplification respectively (see text). Finally, the squares indicate the parameters used in the full wave simulation later.}
	\label{Fig:G-B} 
\end{figure*}

We now investigate the behavior of the same system in the presence of the mirror. We first do so by using the temporal coupled mode theory:
\begin{equation} \label{Eq:EPCMT}
	\begin{aligned}
		\frac{da_\text{cw}}{dt}&=[i(\omega_0-\omega)-2\gamma-\alpha+g]a_\text{cw}+\sqrt{2\gamma} s_1, \\
		\frac{da_\text{ccw}}{dt}&=[i(\omega_0-\omega)-2\gamma-\alpha+g]a_\text{ccw}+\sqrt{2\gamma} s_3 \cdot r e^{i \phi}, \\
		s_3 &=-\sqrt{2\gamma} a_\text{cw},\\
		s_5 &=-\sqrt{2\gamma} a_\text{ccw},
	\end{aligned}
\end{equation}
where we consider $s_5$ to be the output port, $a_\text{cw}$ and $a_\text{ccw}$ are the amplitude of the resonator mode in CW and CCW direction, $r$ is the magnitude of the field reflection coefficient of the mirror, and $\exp(i \phi)$ is an additional phase due to reflection and propagation in the waveguide. In the absence of any input signal, the above system is described by the following effective coupled mode equations:
\begin{equation} \label{H}
	i\frac{d}{dt} \begin{bmatrix}
		a_\text{cw} \\ a_\text{ccw}
	\end{bmatrix}
	=H \begin{bmatrix}
		a_\text{cw} \\ a_\text{ccw}
	\end{bmatrix},
	H=
	\begin{bmatrix}
		\Omega   & 0  \\
		-2i \gamma r e^{i \phi}    & \Omega
	\end{bmatrix},
\end{equation}
where $\Omega=\omega-\omega_0-i(2 \gamma +\alpha -g)$. Interestingly, $H$ is a non-diagonalizable Jordan matrix that features a chiral EP \cite{Zhong2019PRL}, which has also been implemented using nanoscatterers arrangements \cite{Wiersig2014PRL,Wiersig2016PRA}.

Under external driving from port $s_1$, the scattering coefficient between input and output ports is: 
\begin{equation}\label{Eq:s5}
	s_{51} \equiv\frac{s_5}{s_1}=\frac{4r e^{i \phi} \gamma^2}{[i(\omega-\omega_0)+2\gamma+\alpha-g]^2}.
\end{equation}
This solution is valid only below the lasing threshold $g=2\gamma+\alpha$.
Importantly, the scattering coefficient $s_{51}$ exhibits a double pole as compared to the single pole in Eq. (\ref{Eq:s3}). As we will see shortly, this will have drastic consequences. Under these conditions, the maximum value of the amplification is $G_\text{EP} \equiv |s_{51}(\omega_0)|^2=16r^2\gamma^4/(2\gamma+\alpha-g)^4$. On the other hand, the bandwidth  is given by $B_\text{EP}=2F(2\gamma+\alpha-g)$ with $F=\sqrt{\sqrt{2}-1}\approx0.64$. The subscript EP here emphasizes that these quantities are obtained when the system operates at a chiral EP. When comparing these results with those obtained for the DP-based amplifiers, we find that the bandwidth in the current scenario is reduced by a factor of $F$, while the gain is enhanced according to the quadratic relation $G_\text{EP}=r^2 G_\text{DP}^2$. This leads to:
\begin{equation}\label{Eq:GBW}
	\chi_\text{EP} \equiv G_{\text{EP}}^{1/4} \cdot B_{\text{EP}}=4F \sqrt{r} \gamma.
\end{equation}

Equation (\ref{Eq:GBW}) is one of the central results of this work. It shows that the gain-bandwidth product for the EP regime scales differently than for the case of DP. As we will demonstrate below, this provides superior performance over the standard amplifier operating at DP. To facilitate the comparison between the two scenarios (EP vs DP), we set $r \approx 1$, which can be achieved in realistic implementations.

We first consider the case when the two amplifiers based on EP and DP respectively have the same bandwidth. This occurs for different levels of pumpings  related by $\tilde{g}_\text{EP}=F^{-1}\tilde{g}_\text{DP}+2(1-F^{-1})$, where $\tilde{g}=(g-\alpha)/\gamma$, and $\alpha$, $\gamma$ are identical for both amplifiers but $g$ is different. Under these conditions, the amplification enhancement factor $\eta_G$ is: 
\begin{equation}\label{Eq:GEP-B}
	\eta_G \equiv \frac{G_\text{EP}}{G_\text{DP}}=\frac{4F^4}{(2-\tilde{g}_\text{DP})^2}.
\end{equation}
The amplification enhancement for identical bandwidth is plotted in Fig. \ref{Fig:G-B}(a),  together with the pumping relation to achieve identical bandwidth (inset). Above point $p$ (where $\tilde{g}_\text{DP}=2\sqrt{2}F^2\approx1.17$), we have $\eta_G>1$, i.e., the EP-based amplifier outperform the DP one. Notably, the value of the amplification enhancement factor increases rapidly as the two amplifiers approach the lasing condition at $\tilde{g}_\text{DP}=\tilde{g}_\text{EP}=2$.

Next, we consider the situation where both amplifiers have the same maximum amplification ($G_\text{EP}=G_\text{DP}$) but different bandwidth. This condition can be met if $\tilde{g}_\text{EP}=2-\sqrt{2(2-\tilde{g}_\text{DP})}$. In this case, the bandwidth enhancement factor $\eta_B$ is given by:
\begin{equation}\label{Eq:BEP-G}
	\eta_B \equiv \frac{B_\text{EP}}{B_\text{DP}}=\frac{\sqrt{2}F}{\sqrt{2-\tilde{g}_\text{DP}}}.
\end{equation}
Figure \ref{Fig:G-B}(b) depicts $\eta_B$ for increasing power amplification, together with the pumping relations as a function of $\tilde{g}_{\text{DP}}$. As before, the critical point $p$ (identical to that of Fig. \ref{Fig:G-B}(a)) divides the operation domain into two regimes with $\eta_B<1$ (dashed line) and $\eta_B>1$ (solid line). Similar to the previous case, the value of $\eta_B$ increases rapidly (eventually diverging) close to the lasing condition $\tilde{g}_\text{DP}=\tilde{g}_\text{EP}=2$ (not shown in the figure). 

Our discussion clearly demonstrates that operating at an EP can provide superior performance with very large values of $\eta_G$ or $\eta_B$. However, from a practical perspective, the operating point should be chosen sufficiently away from the lasing threshold to avoid noise induced instabilities that can force the system into the lasing regime. Based on the detailed implementation and noise level, this can pose an upper limit on the enhancement factors.

\section{Implementation and full-wave analysis}

\begin{figure}[h]
	\centering
	\includegraphics[width=2.6in]{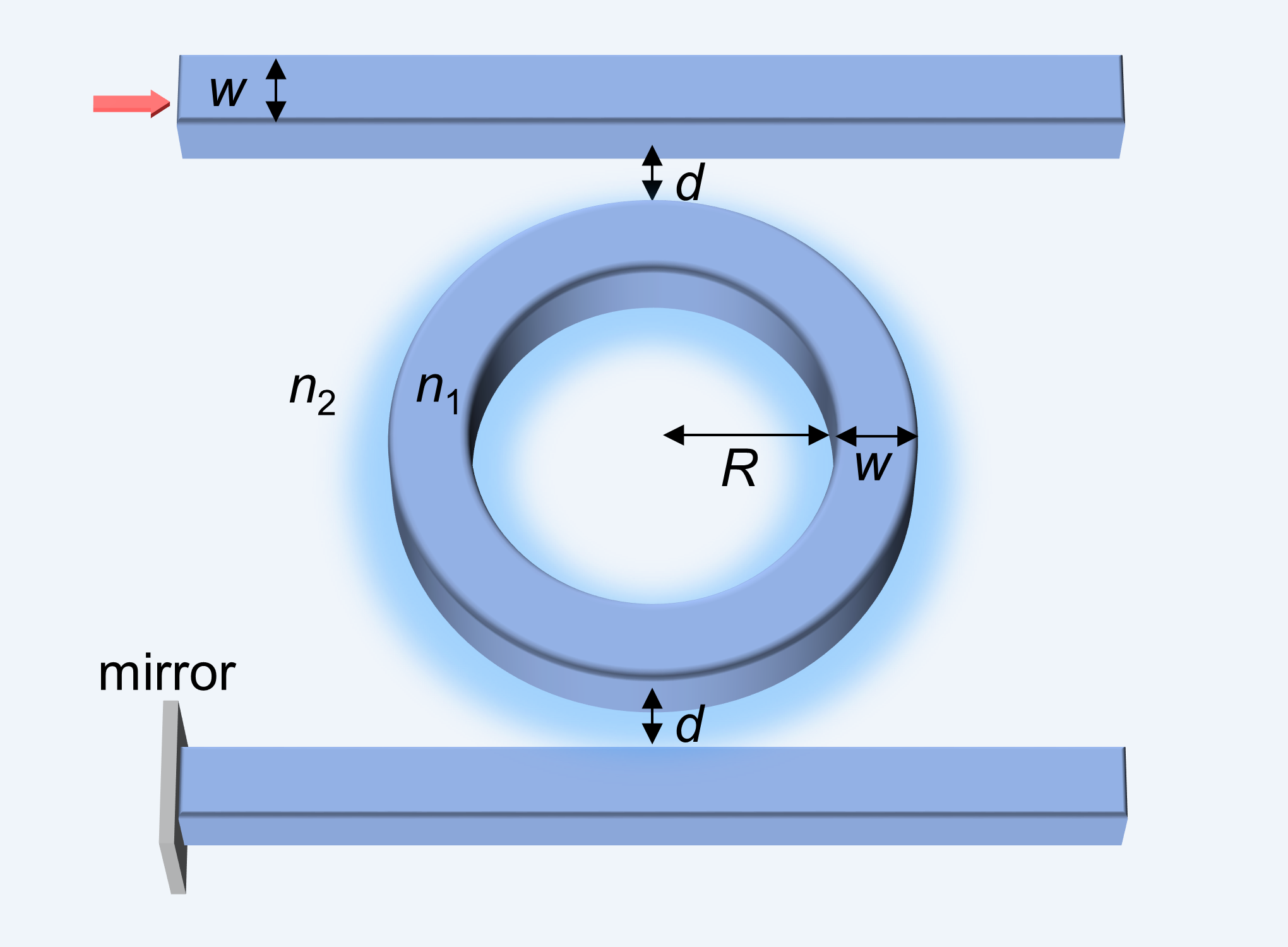}
	\caption{A schematic diagram of the proposed photonic structure used in our FDTD simulations. The geometric and material parameters are assumed to be the following: waveguide width $w$ = 0.25 $\mu$m (for both the straight and the ring waveguides), ring radius $R$ = 5 $\mu$m, edge-to-edge distances between the ring and waveguides $d$ = 0.15 $\mu$m. To implement the mirror, we assume a thin layer of silver with a thickness of 100 nm (In practice, there are several design principles for building a mirror, for instance by using photonic crystals \cite{Loncar2009APL,Loncar2010APL,Loncar2013NL}  or a section of different material \cite{Zamek2011OE}.) The material refractive index is $n_1=3.47$ (corresponding to semiconductor materials such as silicon or AlGaAs) and the background index is taken to be $n_2=1.44$. These values have been used before in DP-based microring amplifiers \cite{Amarnath2005PTL,Amarnath2006}. Finally, we model the applied gain by considering a gain curve with a finite bandwidth (see Appendix A for more details).}
	\label{Fig:Diemnsions} 
\end{figure}

\begin{figure*}[htbp]
	\centering
	\includegraphics[width=5.35in]{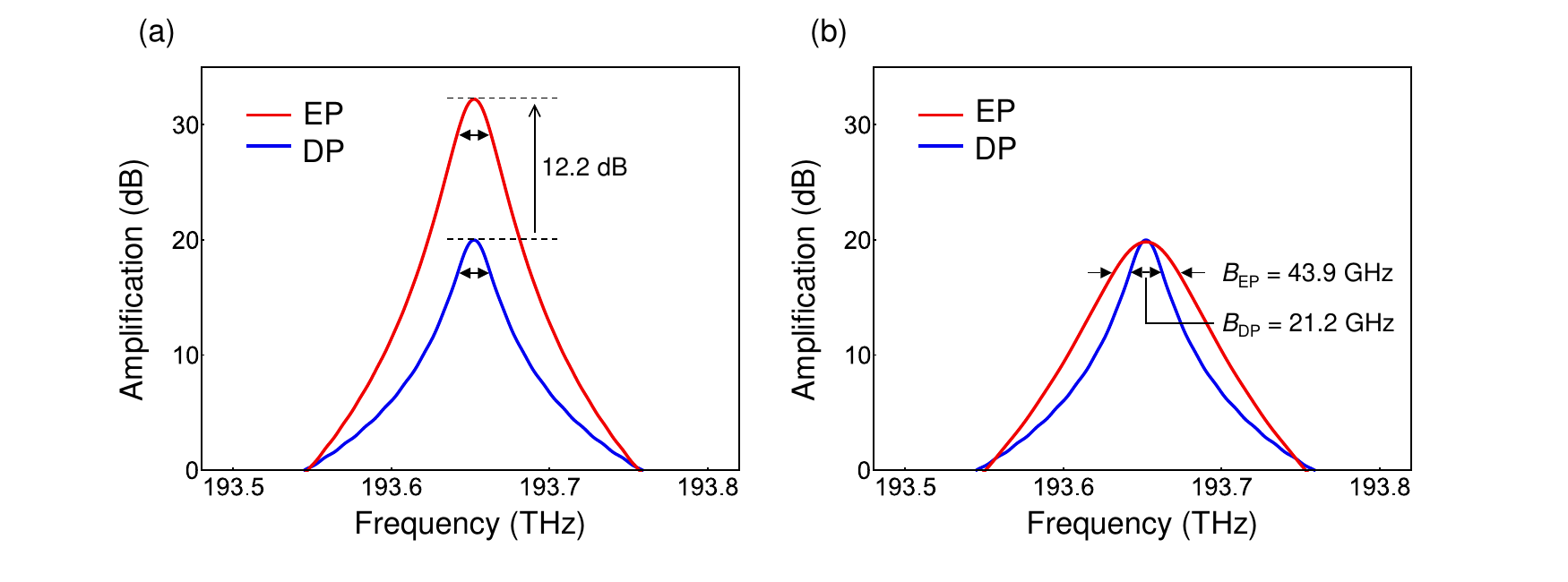}
	\caption{Full-wave FDTD simulations for EP and DP-based amplifiers operating close to $\lambda$ = 1.55 $\mu$m and having (a) identical bandwidth; and (b) identical maximum amplification respectively. The superior performance  larger amplification in (a) and bandwidth in (b) is evident in both cases. The operating points of both scenarios correspond to the square dots in Figs. \ref{Fig:G-B}(a) and (b) correspondingly. Excellent agreement is between the FDTD results and the coupled mode theory is observed in both cases. The details of the design parameters used in our simulations are listed in the text.}
	\label{Fig:FDTD} 
\end{figure*}

\begin{figure}[htbp]
	\centering
	\includegraphics[width=3.4in]{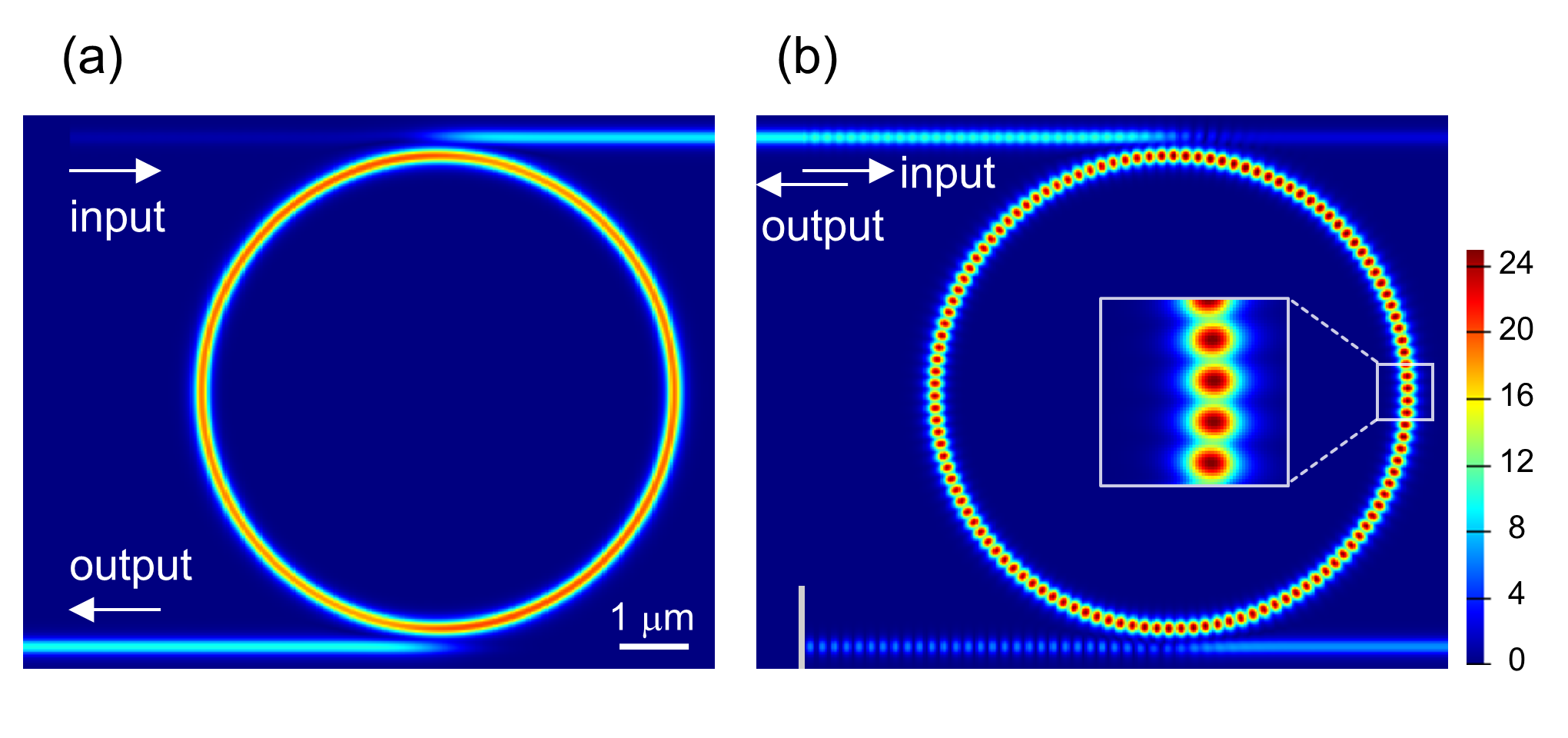}
	\caption{Electric field distributions associated with: (a) DP-based, and  (b) and EP-based amplifier for the resonant frequency when they both have equal gain (i.e. corresponding to the case of Fig. \ref{Fig:FDTD} (b)). The inset in (b) highlights the interference pattern between the CW and CCW components in the latter case. The legend colors represent the value if the electric field normalized by the value of the input field.}
	\label{Fig:EField} 
\end{figure}

We have so far discussed the operation of our proposed EP-based OAs based on the optical coupled mode theory. In order to confirm these predictions, we explore realistic implementations by performing two-dimensional (2D) full-wave FDTD simulations \cite{Taflove-CE}. Particularly, we study a 2D version of the
schematic shown in Fig. \ref{Fig:Diemnsions}. The geometric and physical parameters are all listed in the figure caption.   

Figures \ref{Fig:FDTD}(a) and (b) depict the simulation results for the two different scenarios discussed above, i.e. equal bandwidth or equal maximum amplification. The tuning of the amplifiers to operate in either of these regimes is done by setting up the correct gain parameters (see Appendix A for more details). As shown in Fig. \ref{Fig:FDTD}(a), for fixed bandwidth of 21.2 GHz, a large amplification enhancement factor is achieved with $\eta_G=16.7$, corresponding to 12.2 dB.  On the other hand, Fig. \ref{Fig:FDTD}(b) shows that for an equal maximum amplification of 20 dB, the bandwidth in the EP case can be doubled, $\eta_B=2.1$. These results, which have been obtained by using full-wave FDTD, are consistent with theoretical values predicted by coupled mode theory and clearly indicate the potential utility of the proposed structure.

Finally, Fig. \ref{Fig:EField} plots the field distribution for the two cases of DP and EP amplifiers (corresponding to the structure of Fig. \ref{Fig:Diemnsions} without and with the mirror) for the scenario depicted in Fig. \ref{Fig:FDTD}(b) at resonance. In the EP case, one can observe the interference pattern that results in due to the coexistence of CW and CCW waves. Note that minimum of the field (see inset) is not zero, which can be understood by recalling that the CCW component has larger amplitude (due to amplification) than the CW component. This can be also confirmed by inspecting the time evolution of the fields (not shown here).

\section{Discussions}
The scaling features of the gain-bandwidth products associated with the geometry shown in Fig. \ref{Fig:Setup} can be understood intuitively by noting that light traverses the ring twice in the forward and backward directions, which explains gain enhancement. This observation raises the question of whether one can achieve the same performance by concatenating two ring resonators. As we discuss in Appendix B, this is indeed possible and gives exactly the same results. Interestingly, even in this latter case, one can show that the system exhibits a second-order EP, though an unusual one (see Appendix B for detailed discussion). This provides an advantage in terms of scalability since one can add more microrings to obtain even higher order poles with far more superior performance. However, from a practical perspective, this latter design (with concatenated rings)  will be more prone to fabrication errors (all the different ring parameters have to exactly match) and will require more complex pumping scheme. This is in contrast to the design of Fig. \ref{Fig:Setup} which does not suffer from these problems. 

Another possible implementation that can combine the enhanced performance with the robustness is the S-bend ring resonator, which is also known to provide unidirectional coupling \cite{Hohimer1993APL,Kharitonov2015LSA,Khajavikhan2018OE}. As shown in Appendix C, the output power demonstrates similar scaling behavior with that of EP-based OAs with a mirror.

In conclusion, we have introduced a new design paradigm for optical amplifiers based on chiral exceptional points. An important feature of the proposed structure is the unique scaling of its gain-bandwidth product which is different from standard amplifiers, and allows for achieving more gain or larger bandwidth of operation. Mathematically, these results can be understood by noting that operating at an EP results in a double pole in the scattering coefficients (as opposed to a single pole in the standard DP case). Importantly, we have explored realistic implementations using current photonics technology to implement these amplifiers based on chiral exceptional points (for completeness, we have also confirmed these conclusions for parity-time-(PT-)based EP in Appendix D). Our results open the door for building a new generation of on-chip optical amplifiers superior performance, which can prove beneficial for both classical and quantum optics applications. Finally, we emphasize that our proposed scheme can be directly mapped into other physical systems such as microwaves and acoustics.

\appendix
\section{FDTD simulation}

\begin{figure}[bp]
	\centering
	\includegraphics[width=2.6in]{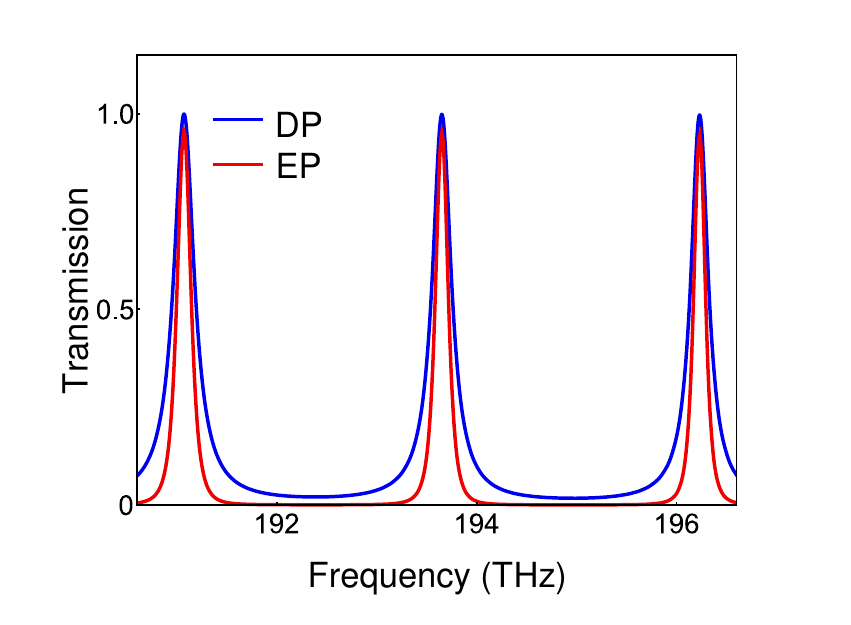}
	\caption{The transmission of the DP-based and EP-based resonators without any material gain. Then a material gain based on Lorentz model as discussed in the text was applied to the microring resonator.}
	\label{Fig:FDTDApp} 
\end{figure}

The transmission of the passive resonators used in our simulations based on DP or EP without any material gain is shown in Fig. \ref{Fig:FDTDApp}. The operating resonant frequency is located at $f_0=193.652$ THz, with free spectral range (FSR) of 2.58 THz. The maximum transmission for the DP-based resonator is 0.998, indicating $\alpha/\gamma=0.002$, which is a small quantity. The maximum transmission for the EP-based resonator is 0.952, which can be used to deduce the value of reflection coefficient $r^2=0.954$ --- consistent with the $r^2=0.953$ obtained from a direct FDTD simulation test on reflectivity of the mirror. 

In our simulations, the applied material gain has a finite bandwidth as expressed by the Lorentz model: 
\begin{equation}
\varepsilon(\omega)=\varepsilon_\text{b}+\frac{\varepsilon'\omega_0^2}{\omega_0^2-\omega^2-2 i \delta \omega},
\end{equation}
where $\varepsilon_\text{b}$ is the permittivity of background material in the absence of any gain/loss or dispersion; $\omega_0$ is operating resonant frequency of the microring; $\delta=10^{13}$ rad/s is gain curve linewidth; $\varepsilon'$ is a constant. To proceed with the computations, we set the value of $\varepsilon'$ for every case and use FDTD to calculate the maximum amplification at resonant. This quantity can be then used to obtain the normalized gain values $\tilde{g}_{\text{DP}}$ and $\tilde{g}_{\text{EP}}$ (see the formulas for $G_{\text{DP}}$ and $G_{\text{EP}}$), which in turn allows us to compare our FDTD results with Eqs. (6) and (7) in the main text.  Particularly, in the simulations of Fig. 3 in the main text, we used $\varepsilon'=-2.133\times10^{-4}$ for the DP amplifier;  $\varepsilon'=-2.00194\times10^{-4}$ and $\varepsilon'=-1.6198\times10^{-4}$ for the EP amplifiers of Figs. 3(a) and (b), respectively.

\section{Cascaded amplifiers and exceptional points}
\begin{figure}[htbp]
	\centering
	\includegraphics[width=2.6in]{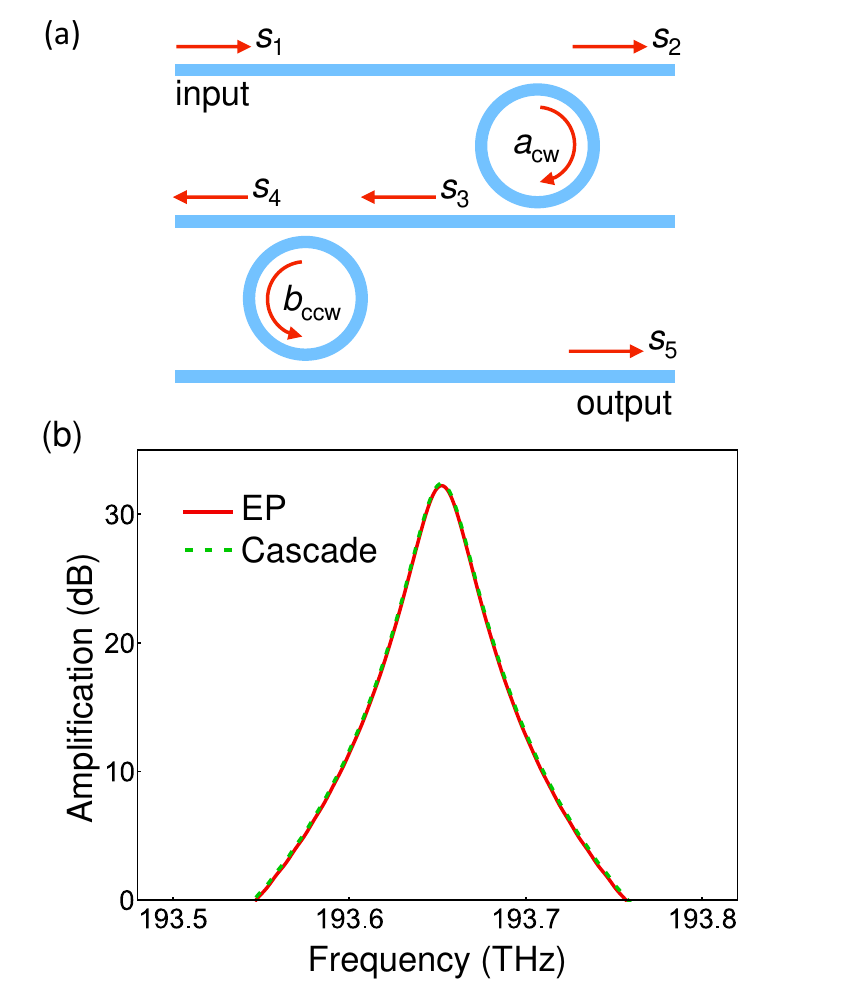}
	\caption{(a) A cascaded amplifier  can achieve the same functionality as the structure in Fig. 1 with $r\exp(i \phi)=1$, as confirmed in (b) using FDTD.}
	\label{Fig:Cascade} 
\end{figure}

In the EP-based amplifier proposed in Fig. 1, light travels from port $s_1$ to $s_3$, gets reflected and travels back to the same input port in the opposite direction. This intuitive picture can explain the enhacement in the net amplification. It also raises the question of whether it is possible to achieve the same functionality by concatenating two ring resonators. By referring to Fig. \ref{Fig:Cascade}(a), it is not difficult to see that this structure has identical scattering coefficient to that of Fig. 1 with  $r \exp(i\phi)=1$. This is also confirmed by using FDTD with $\varepsilon'=-2.00194\times10^{-4}$, as shown in Fig. \ref{Fig:Cascade}(b). 

At first sight, this may seem surprising but interestingly, the system in Fig. \ref{Fig:Cascade} also has a chiral EP since mode $a_\text{cw}$ couples to $b_\text{ccw}$ while the opposite is not true. In fact, here there is also another chiral EP arising from the unidirectional coupling from $b_\text{cw}$ to $a_\text{ccw}$, which would allow the amplifier to work for backward propagating light as well. Interestingly, this scheme can be used to build amplifiers with higher order EPs by just cascading as many rings as needed, thus provides a clear advantage in terms of scalability. In practice, however, this system will be more prone to fabrication errors since it will require all the rings to have identical parameters within a small margin of error (disorder in the coupling parameters will not affect the chiral EP). Additionally, it will also require a complex pumping scheme and more power consumption to provide gain to all the rings. On the other hand, the structure proposed in Fig. 1 does not suffer from these problems. Particularly, any variation in the ring parameter will affect both modes equally which will shift the central frequency but retain the same gain-bandwidth relation. Additionally, it contains only one ring and thus requires simpler pumping and less power consumption.

\section{S-bend ring resonator}

\begin{figure}[htbp]
	\centering
	\includegraphics[width=2.6in]{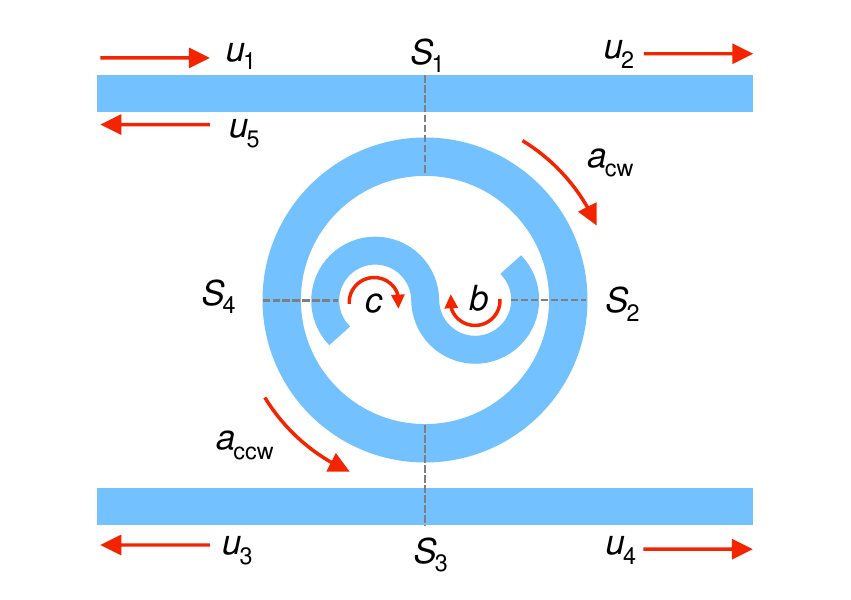}
	\caption{The S-bend ring can provide a unidirectional coupling between CW and CCW mode. This structure is studied with scattering matrices $S_j$ ($j=1,2,3,4$) in the four coupling regions (dashed lines).}
	\label{Fig:S-Bend} 
\end{figure}

Another possible implementation that can combine the enhanced performance with the robustness is the S-bend ring resonator shown in Fig. \ref{Fig:S-Bend}. This structure is also known to provide unidirectional coupling \cite{Hohimer1993APL,Kharitonov2015LSA,Khajavikhan2018OE}. By using the scattering matrices $S_j=\begin{bmatrix} t_j & i \kappa_j \\ i \kappa_j & t_j\end{bmatrix}$ (with $t_j^2+\kappa_j^2=1$, where $j=1,2,3,4$) at each junction (denoted by the dashed black lines in Fig. \ref{Fig:S-Bend}), we obtain the relation between the electric field amplitudes $a_\text{cw}$ and $a_\text{ccw}$ at the beginning and end of each section along the ring between any two junctions. This, in turn, can be used to calculate the scattering coefficients. Particularly, when $S_2=S_4$ and $S_3=I$, where $I$ is the unit matrix (i.e. remove the lower waveguide altogether), we obtain
\begin{equation}
u_{51}\equiv\frac{u_5}{u_1}=\frac{2\Gamma t_2 \kappa_1^2 \kappa_2^2 \exp(i \omega \tau)}{[1-\Gamma t_1 t_2^2 \exp(i \omega \tau)]^2},
\end{equation} 
where $\Gamma=\exp[-2\pi \text{Im}(n_{\text{eff}}) L /\lambda]$ is the round trip gain of ring resonator, and $\tau=\text{Re}(n_\text{eff}) L/c$. Here $n_\text{eff}$ is the effective index of ring waveguide mode, $L$ is the circumference of the ring waveguide and $\lambda$ is the free space wavelength. The maximum amplification (at resonant frequency) and the bandwidth are then found to be:
\begin{equation}
\begin{aligned}
G_\text{S}&=\frac{4 \Gamma^2 t_2^2 \kappa_1^4 \kappa_2^4}{(1-\Gamma t_1 t_2^2)^4}, \\
B_\text{S}&=2F \frac{1-\Gamma t_1 t_2^2}{\sqrt{\Gamma t_1} t_2} \tau^{-1}.
\end{aligned}
\end{equation}
Consequently, the gain-bandwidth product is given by:
\begin{equation}
\chi_\text{S}\equiv G_\text{S}^{1/4} \cdot B_\text{S}=2\sqrt{2}F\kappa_1\kappa_2 t_1^{-1/2} \tau^{-1},
\end{equation}
with the right hand side being a constant. This last expression reveals that the gain-bandwidth product in the S-bend geometry scales in a similar fashion to the structure shown in Fig. 1. From an experimental perspective, one can add the second waveguide and measure the output from $u_4$, where it can be shown that the output power demonstrates similar scaling behavior.  

\section{Amplifiers at EPs in PT symmetric dimers}

\begin{figure}[htbp]
	\centering
	\includegraphics[width=2.6in]{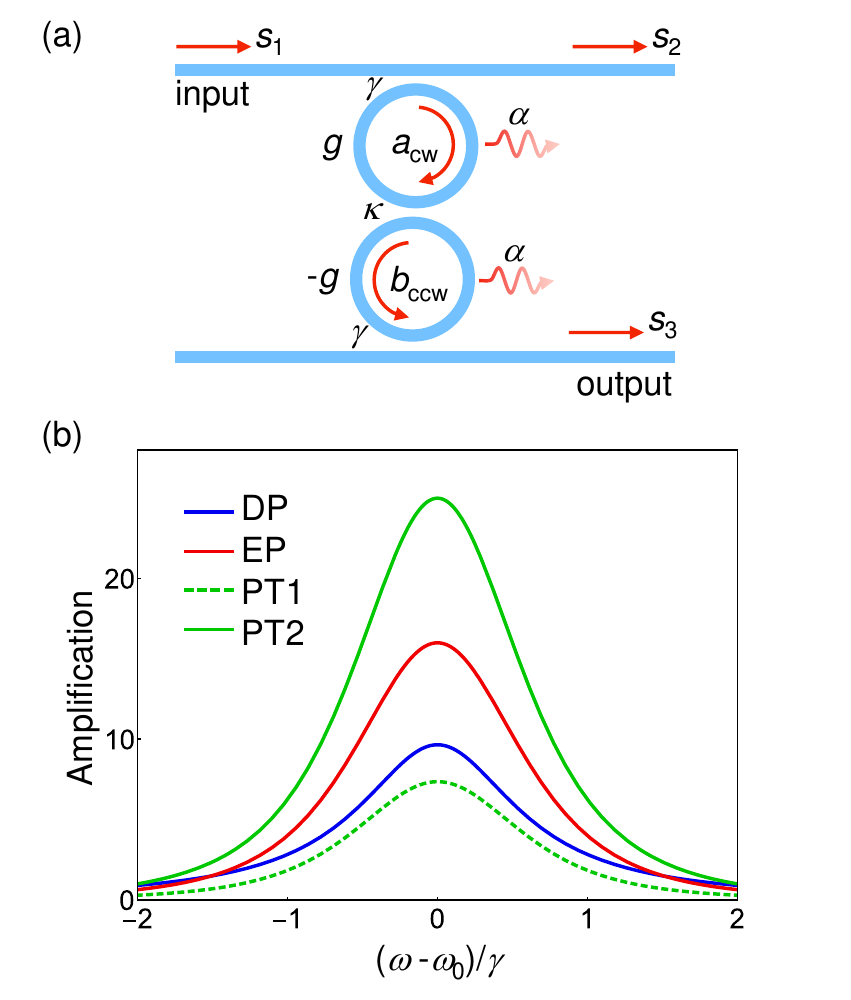}
	\caption{(a) A PT dimer based optical amplifier having an exceptional point at $g=\kappa$. (b) Comparison between the strcuture in (a) and those in Fig. 1 (with and without a mirror) for identical $\gamma$ and assuming $\alpha=0$, when $g_\text{PT1}=1.36\gamma$ (dashed green line) and $g_\text{PT2}=2.5\gamma$ (solid green line). Clearly, one can increase the amplification of the PT amplifier while at the same time maintain the same bandwidth. From a practical perspective however, increasing the amplification requires stronger coupling $\kappa$ (i.e. smaller separation between the two rings) which is limited by the fabrication tolerance.}
	\label{Fig:PT} 
\end{figure}

In order to make the connection between our results here and the work on PT symmetry; and at the same time illustrate that the predicted superior performance of EP-based amplifiers is general and not restricted to the geometry investigated in the main text, we consider an amplifier based on the archetypal PT symmetric dimer \cite{El-Ganainy2007OL,Guo2009PRL,El-Ganainy2010NP,Khajavikhan2014S,Yang2014NP,Chang2014NP,Schindler2011PRA,El-Ganainy2018NP,Feng2017NP}  shown in Fig.  \ref{Fig:PT}(a). It consists of two identical microrings coupling with each other with coupling rate $\kappa$. Both rings have the same radiation loss and coupled to identical waveguides with equal coupling coefficients. Additionally, we assume that the top ring has a material gain $g$ while the lower ring has an additional loss factor $-g$. By using TCMT, we obtain:
\begin{equation} \label{Eq:PTCMT}
\begin{aligned}
\frac{da_\text{cw}}{dt}&=[i(\omega_0-\omega)-\gamma-\alpha+g]a_\text{cw}+i \kappa b_\text{ccw}+\sqrt{2\gamma} s_1, \\
\frac{db_\text{ccw}}{dt}&=[i(\omega_0-\omega)-\gamma-\alpha-g]a_\text{ccw}+i \kappa a_\text{cw}, \\
s_3 &=-\sqrt{2\gamma} b_\text{ccw}.
\end{aligned}
\end{equation}
By solving the above system, we obtain the expression for the steady state transfer function: 
\begin{equation}\label{Eq:PT-s3}
s_{31} \equiv \frac{s_3}{s_1}=\frac{-2ig\gamma}{i[(\omega-\omega_0)+\gamma+\alpha]^2+\kappa^2-g^2}.
\end{equation}

Note that this solution is valid only below the lasing threshold $g^2=\kappa^2+(\gamma+\alpha)^2$. The system exhibits an EP when $g=\kappa$. Under this condition, the maximum amplification is given by $G_\text{PT}=4g^2 \gamma^2 / (\gamma+\alpha)^4$ while the bandwidth take the value  $B_\text{PT}=2F(\gamma+\alpha)$. Interestingly, in contrasts to the structure investigated in the main text, here the bandwidth is independent of the gain. In other words, the gain-bandwidth product can be made arbitrary large by judicious choice of the design parameters and the applied gain --- a feature that was previously noted for linear microwave amplifiers based on wave mixing processes \cite{Metelmann2014PRL,Roy2015APL}, though without establishing the connection with the physics of exceptional points. To illustrate this point, Figure \ref{Fig:PT}(b) depicts a comparison between the three different structures of Fig. 1 (with and without a mirror) and Fig. \ref{Fig:PT}(a). Here we choose $\alpha \approx 0$ and identical $\gamma$ for all three devices. The bandwidth of PT-based amplifier is $2F\gamma\approx1.28\gamma$. This same bandwidth can be achieved for the other two geometries by setting $g_\text{DP}=(2-F)\gamma\approx1.36\gamma$ and $g_\text{EP}=\gamma$. When $g_\text{PT}=g_\text{DP}$, the PT system exhibits lower amplification as shown by the dashed green line in Fig. \ref{Fig:PT}(b). However, in theory, the maximum amplification can be increased indefinitely by increasing  $g_\text{PT}$ without affecting the bandwidth. For example, by choosing  $g_\text{PT}=2.5\gamma$, the maximum amplification of the PT structure can significantly surpass that of the other two scenarios while at the same time maintaining the same bandwidth (solid green line in Fig. \ref{Fig:PT}(b)).

From a practical perspective however, increasing the gain bandwidth product will require fabricating samples with stronger coupling between the two rings which is obviously limited by the minimum achievable edge-to-edge separation between the rings. Additionally, the PT geometry is very sensitive to fabrication errors and tolerance as well as uncertainties in the operating conditions such as thermal effects for instance.\\

\begin{acknowledgments}
R.E. and S.K.O. acknowledge support from the National Science Foundation (Grant No. ECCS 1807552). A.M. is supported by Emmy Noether program (Grant No. ME 4863/1-1), Deutsche Forschungsgemeinschaft. A.E. acknowledges support from the Deutsche Forschungsgemeinschaft via a Heisenberg fellowship (Grant No. EI 872/5-1).  
\end{acknowledgments}

\end{document}